\documentclass[dvips,11pt]{article}
\usepackage{varioref,exscale,latexsym,amsmath,amssymb}
\usepackage{graphicx}

\def\beq{\begin{equation}}

\def\eeq{\end{equation}}

\def\beqa{\begin{eqnarray}}

\def\eeqa{\end{eqnarray}}

\title{{\bf Long distance chiral corrections in B meson amplitudes}}
\author{Juan J. Sanz-Cillero $^a$\\
John F. Donoghue$^b$, and Andreas Ross$^{b,c}$\\ \\
$^a$ Departament de F\'\i sica Te\`orica, IFIC,\\
Universitat de Val\`encia - CSIC\\
Valencia, Spain\\  \\
$^b$ Department of Physics\\
University of Massachusetts\\
Amherst, MA  01003\\  \\
$^c$ Institut f\"ur Theoretische Teilchenphysik\\
Universit\"at Karlsruhe\\
Karlsruhe, Germany}
\begin{document}
\begin{titlepage}
\maketitle
\begin{abstract}
We discuss the chiral corrections to $f_B$ and $B_B$ with particular emphasis on
determining the portion of
the correction that arises from long distance physics.
For very small pion and kaon masses all of the usual corrections are truly long
distance, while for larger masses the long distance portion decreases. These chiral
corrections have been used to extrapolate lattice calculations towards the physical
region of lighter masses. We show in particular that the chiral extrapolation is
better behaved if only the long distance portion of the correction is used.
\end{abstract}
\vspace{0.5 in}
IFIC/03-09 \\
FTUV/03-0409 \setcounter{page}{0}
\end{titlepage}
\section{Introduction}
Lattice calculations of B meson properties are presently done with parameters such
that the light quark masses are larger than their physical values. In order to make
predictions that are relevant for phenomenology, these calculations are extrapolated
down to lower quark masses. One of the extrapolation methods uses some results from
chiral perturbation theory, and this appears to produce rather large effects due to
the chiral corrections. A recent summary of the field~\cite{lellouch} noted that
this chiral extrapolation is the largest uncertainty (17\%) at present in the
calculation of the B meson decay constant $f_B$.

 Chiral perturbation theory is an effective field theory involving pions,
kaons and $\eta$ mesons. These mesons are the lightest excitations in QCD and the
effective field theory is designed to describe the effects of long range
propagation of these light degrees of freedom. Even in loop diagrams there are long
distance effects which are described well by the effective field theory. However,
chiral perturbation theory is not a good model of physics at short distances and is
not valid for large meson masses. If we consider mesons of variable mass, as the
masses become heavier, less and less of the loop corrections are truly long distance.

The chiral corrections are sometimes used in ways that hide the separation of long
distance and short distance physics. Consider for example the chiral correction to
the B meson decay constant in dimensional regularization~\cite{grinstein, sharpe,
Becirevic:2002sc}
\begin{equation}
f_B  =  f_0 \left[1 - \left({1+3g^2
  \over 16\pi^2 F_\pi^2}\right) \, { 3 \over 8} \,
 m_\pi^2 \ln {m_\pi^2 \over\mu^2} + ... \right] \, ,
\end{equation}
where $g$ is the heavy meson coupling to pions. The ellipses denote the kaon and eta
contributions as well as analytic terms in the masses that carry unknown
coefficients which must be fit. We see that the corrections vanish for massless
mesons and grow continuously with large meson masses\footnote{Note that we keep the
B meson mass unchanged, so that when we refer to large and small meson masses, we
are always referring to the masses of the chiral particles - pions, kaons and etas -
that occur in the loop diagrams. }. This is the opposite of the behavior that one
might expect, which would be to have larger chiral corrections when the pions are
nearly massless. For very large masses of the ``pions'', physically we expect that
the loop effects must decouple from the observables. The expression of Eq. 1 does
not illustrate this decoupling. The key point is that as the mesons become heavier,
most of the correction given in Eq. 1 comes from short distance physics, which is
not a reliable part of the effective field theory. We will show this in more detail
below. This behavior is not a problem in principle. The free coefficients in the
chiral lagrangian allow one to compensate for the unwanted behavior and correctly
match the short distance physics of QCD. However the reliance on Eq. 1 at large
masses can have a deleterious effect on phenomenology in some applications.

The way that present lattice extrapolations of $F_B$ are performed apply the chiral
predictions outside their region of validity. An example is given in Fig. 1,
describing the results
of the JLQCD collaboration~\cite{JLQCD}.\\
\begin{figure}[h]
\begin{center}
  \begin{minipage}[t]{.07\textwidth}
    \vspace{0pt}
    \centering
    \vspace*{122pt}
    \hspace*{-10pt}
    \rotatebox{90}{$f_B$}
  \end{minipage}%
  \begin{minipage}[t]{0.93\textwidth}
    \vspace{0pt}
    \centering
    \includegraphics[width=0.99\textwidth,height=!]{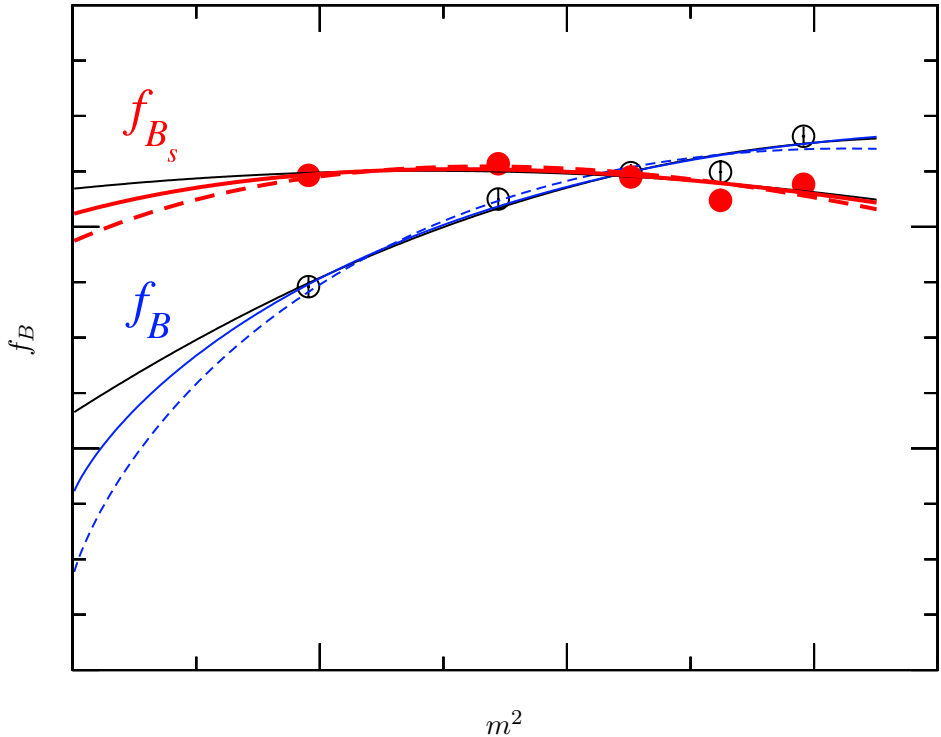}
  \end{minipage}
\end{center}
$\hspace*{185pt} m^2$
\caption{Lattice data points for $f_B$ and $f_{B_s}$ and fitted curves with quadratic
fit (upper solid curve) and with chiral
         logs for $g=0.27$ and $g=0.59$ (dashed)\vspace*{10pt}}
\end{figure}
${}$\\
In order to address the issue of the chiral exptrapolation, the lattice data was fit
with the function of Eq. 1 at large mass and the form is used to extrapolate the
results to small values of the mass. The fact that there appears to be a large
effect at $m=0$ does not imply that the chiral correction is large here. Indeed,
inspection of Eq. 1 shows that the chiral log correction vanishes at zero mass, so
the chiral logarithm is not large at the physical masses. Rather, the big effect
seen comes from using Eq. 1 at large masses. Since the chiral logs grow at large
mass, and appear in this formula with a fixed coefficient, normalizing the function
at large mass produces a sizeable difference when compared to smaller masses. Since
chiral perturbation theory is not applicable at such large masses, this shift is not
a valid consequence of chiral perturbation theory.

We will explore the long-distance/short-distance structure of the chiral corrections
\cite{ldr}, and show that the undesirable effects described above come from short
distance physics that chiral perturbation theory is not able to describe. The
application of Eq. 1 at large masses then amounts to a bad model of the short
distance physics. We will give formulas for the one loop corrections of Eq. 1 which
{\it removes} the unwanted short-distance component. At small quark masses, our
method is just a different regularization of the theory, and reproduces the usual
chiral corrections. When applied at large quark masses, our formulas must also be
considered as a model. However, it is a relatively innocuous model in that it makes
no assumptions about short distance physics and and it produces a small correction
since the loop effect decouples at large mass.
 When used to extrapolate
the lattice results to the physical masses, our results lead to more reasonable
estimates of the chiral corrections. Our methods are similar to some work on long
distance regularization in baryon chiral perturbation theory \cite{ldr} and on
chiral extrapolations in other processes \cite{leinweber}. In particular, the JLQCD
group has explored the use of the Adelaide-MIT approach \cite{leinweber}in the
extrapolation of the pion decay constant\cite{JLQCD}. Our work describes the
rationale and benefits of a modified approach for the heavy-light system.

\section{A study of the chiral corrections to $f_B$}

The chiral corrections were initially calculated by Grinstein et al \cite{grinstein}
(see also \cite{sharpe, becirevic, Becirevic:2002sc}). The methods are standard and
we will not reproduce the details. However we note that, although there are various
Feynman diagrams in the calculation, in the end the loop calculations involve only
one loop integral,
\begin{equation}
\mathbf{I}\left(m\right) =
    i \int \frac {d^4 k} {\left(2 \pi\right)^4}
     \frac {1} {\left(k^2 - m^2 + i \epsilon\right)} \, .
\end{equation}

The chiral expansion involves unknown parameters for the reduced decay constant at
zero mass ($\bar{f}_0$) and for the slopes ($\alpha_1,\alpha_2$) parameterizing
linear dependence in the masses. The results are \cite{grinstein, sharpe,
Becirevic:2002sc}
\begin{align}
 f_{B_{u,d}} = \frac {1} {\sqrt {m_B}} \, & \bar{f}_0\Bigg[1 + \alpha_1 m_\pi^2 +
\alpha_2  (2 m_K^2 + m_\pi^2) \notag\\
 & \hspace*{-2pt} - \, \frac {1+3g^2} {4 \, F_\phi^2} \bigg(\frac {3} {2} \,
   \mathbf{I} \left( m_\pi\right) + \mathbf{I} \left(m_K \right) + \frac
   {1} {6} \, \mathbf{I} \left(m_\eta \right) \! \! \bigg)\Bigg]
\end{align}
and
\begin{align}
 f_{B_s} = \frac {1} {\sqrt {m_B}} \, & \bar{f}_0\Bigg[1 + \alpha_1 (2 m_K^2 - m_\pi^2)
+ \alpha_2 (2 m_K^2 + m_\pi^2) \notag\\
 & \hspace*{-2pt} - \, \frac {1+3g^2} {4 \, F_\phi^2} \bigg( 2 \, \mathbf{I}
 \left(m_K\right) + \frac {2} {3} \, \mathbf{I} \left(m_\eta\right)\bigg)\Bigg]
 \, ,
\end{align}
where $g$ is the coupling of heavy mesons to pions\footnote{In our numerical work,
we will use $g = 0.59$.} and $F_\phi$ is the pseudo-goldstone meson decay constant
in the chiral limit\footnote{We use the normalization such that $F_\pi
=0.0924$~GeV.}. Of course, the integral still needs to be regularized. In
dimensional regularization, one absorbs the $1/(d-4)$ divergences into the slopes
and finds the residual integral
\begin{align}
  \mathbf{I}^{d.r.}\left(m\right) =
    \frac {1} {16 \pi^2} & \Bigg[
       m^2 + m^2 \ln \frac {m^2} {\mu^2} \Bigg] \, ,
\end{align}
where $\mu$ is the arbitrary mass parameter that enters in dimensional
regularization. The physical results do not depend on $\mu$ as it can be absorbed
into a shift in the unknown slope coefficients.

Let us explore the loop integral and study the long-distance part. In order to do
this, we use a cut-off defined in the rest frame of the B meson in order to remove
the short-distance component. Specifically, we use a dipole cutoff yielding
\begin{equation}
\mathbf{I}\left(m, \Lambda\right) =
   i \Lambda^4 \int \frac {d^4 k} {\left(2 \pi\right)^4}
     \frac {1} {\left(k^2 - m^2 + i \epsilon\right) \left(k^2 -
\Lambda^2 + i \epsilon\right)^2} \, .
\end{equation}
In related contexts, other forms of cut-offs have been studied \cite{ldr,leinweber}
- qualitatively similar results are found with other forms, although the parameter
$\Lambda$ will have different meanings in each case. We employ a finite value for
the cut-off of order the size of the B meson. The integral may be calculated and has
the form
\begin{equation}
\mathbf{I}\left(m, \Lambda\right) = \frac { \, \Lambda^4} {16 \pi^2} \Bigg[
     - \frac {1} {m^2-\Lambda^2} + \frac {m^2} {\left(m^2-\Lambda^2\right)^2} \,
     \ln \frac {m^2} {\Lambda^2} \Bigg] \, .
\end{equation}
More illuminatingly, this result is shown in Fig. 2. In this figure we compare the
dimensionally regularized result to the long-distance portion, defined by Eq. 7.

\begin{figure}[h]
\begin{center}
 \begin{minipage}[t]{0.93\textwidth}
    \vspace{-0pt}
    \hspace{-10pt}
    \centering
    \includegraphics[width=0.99\textwidth,height=!]{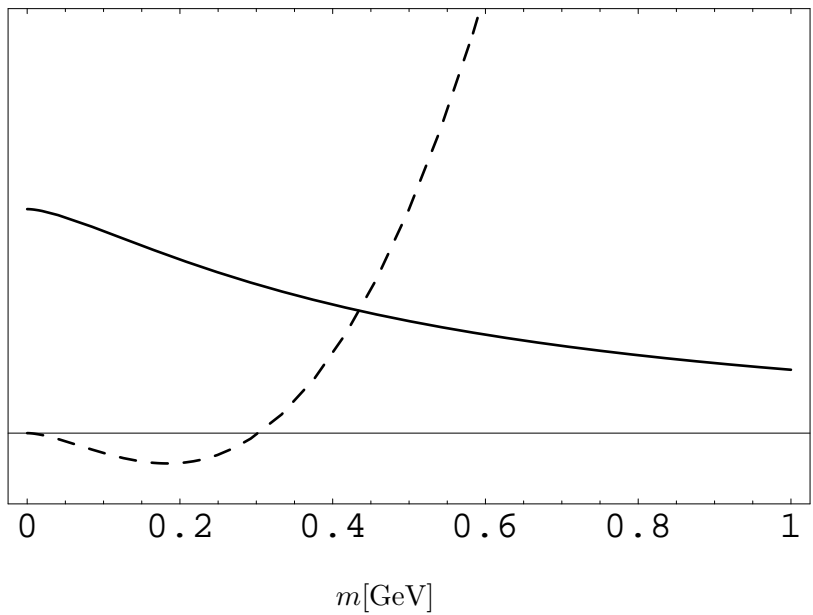}
  \end{minipage}
\end{center}
$\hspace*{149.5pt} m [\mbox{GeV}]$ \caption{Integrals $\mathbf{I}\left(m,
\Lambda\right)$ with $\Lambda = 500$ MeV and $\mathbf{I}^{d.r.}\left(m\right)$
         with $\mu = 500$ MeV (dashed) \vspace*{10pt}}
\end{figure}
The long distance component is seen to have several reassuring features in the
cut-off regularization. It is
largest when the meson is massless, as one would expect. It is small when the mass
is big and exhibits decoupling, vanishing as the mass goes to infinity. It smoothly
interpolates between these limits. When comparing it to the dimensionally
regularized result, one sees a shift in the intercept at zero mass - this is not
surprising because the regularization corresponds to
removing the value when $m=0$.
One also notices that, aside from this shift, both forms
have the same logarithmic behavior near
$m=0$. The small curvature noted at the smallest mass values is the nonlinear
behavior due to the chiral log factor $m^2 \ln m^2$.  Without this term the result
would be able to be Taylor expanded about $m=0$, with the first term being a linear
slope in $m^2$ - the nonlinear behavior is the result of the logarithm.

We also see that the chiral log by itself grows large quickly and has a large
curvature at large masses in dimensional regularization.
This effect is not mirrored in the long-distance
component, so that it is clear that this behavior comes from the short-distance
portion of the integral. This is not surprising. In dimensional regularization,
there is no scale within the integration aside from the particle's mass, so that the
the whole integral scales with $k \sim m$. These short distance effects are ones
which are not reliable calculated by the effective field theory.

These results suggest that we should consider an extrapolation that only includes
the long distance loop effects. The short distance effects are provided by the
lattice simulation\footnote{The "smooth matching" procedure of Ref.
\cite{Becirevic:2002sc} is another attempt to apply the chiral results only in their
region of validity. }. The truly long distance effects are supplied by chiral
perturbation theory. We will use the long distance parts of the loops in performing
the matching of the two regions. In our approach this matching is described by the
parameter $\Lambda$ specifying the separation of long and short distances. The
residual dependence on this parameter, within some range, is a reflection of the
present uncertainty in the matching procedure.

\section{Long distance regularization of the chiral calculation}

At small quark masses, the cut-off treatment of the integral can be promoted to a
regularization of chiral perturbation theory. This has been studied in the context
of baryon chiral perturbation theory in Ref. \cite{ldr}, where it was called long
distance regularization. The use of a cut-off is clearly more painful
calculationally than the usual dimensional regularization, but when the masses are
small it reproduces the usual one-loop chiral expansion for matrix elements such as
we are studying.

In order to regularize the calculation using the cut-off, the divergent pieces
are separated in the Feynman integral. The result is
\begin{align}
  \mathbf{I}\left(m, \Lambda\right) =
    \frac {1} {16 \pi^2} & \Bigg[
      \Lambda^2 - m^2 \, \ln \frac {\Lambda^2} {\mu^2} \Bigg] \,
      + \, \mathbf{I^{ren}}(m,\Lambda) \, ,
\end{align}
where $\mathbf{I^{ren}}(m,\Lambda)$ is finite in the limit $\Lambda\to\infty$. This
residual integral has the form
\begin{align}
  \mathbf{I^{ren}}\left(m, \Lambda\right) =
  \mathbf{I^{d.r.}}\left(m\right) \, + \,
    \frac {1} {16 \pi^2}  \Bigg[  - \frac {m^4} {m^2-\Lambda^2}
        - \frac {m^4 (m^2-2 \Lambda^2)} {(m^2-\Lambda^2)^2} \,
  \ln \frac {m^2} {\Lambda^2}
       \Bigg] \, .
\end{align}

We see that there are potentially divergent contributions proportional to
$\Lambda^2$ and $\ln \Lambda^2$. However, these
have exactly the right structure to be
absorbed into the chiral parameters. In particular, the renormalization is
\begin{align}
   \bar{f}_0^{ren} & = \bar{f}_0 - \frac {8} {3} \ \bar{f}_0 \, \frac {1+3g^2} {64 \pi^2
\, F_\phi^2}  \, \Lambda^2 \notag\\
   \alpha_1^{ren} & = \alpha_1 + \frac {5} {6} \ \, \frac {1+3g^2} {64 \pi^2 \,
F_\phi^2}  \, \ln \frac{\Lambda^2} {\mu^2} \notag\\
   \alpha_2^{ren} & = \alpha_2 + \frac {11} {18} \ \, \frac {1+3g^2} {64 \pi^2 \,
F_\phi^2}  \, \ln \frac{\Lambda^2} {\mu^2}.
\end{align}
After renormalization, we can express the chiral amplitudes in terms of these
parameters plus the logarithmic contribution in the residual integral
$\mathbf{I^{ren}}\left(m, \Lambda\right)$, providing the renormalized
observables
\begin{align}
 f_{B_{u,d}} = &\frac {1} {\sqrt {m_B}} \, \bar{f}_0^{ren} \Bigg[1 +
     \alpha_1^{ren} m_\pi^2 + \alpha_2^{ren} (2 m_K^2 + m_\pi^2) \notag\\
 & \hspace*{-2pt} -   \, \frac {1+3g^2} {4 \, F_\phi^2} \bigg(\frac {3} {2} \,
     \mathbf{I^{ren}} \left(m_\pi, \Lambda\right) + \mathbf{I^{ren}} \left(m_K,
\Lambda\right)
     + \frac {1} {6} \, \mathbf{I^{ren}} \left(m_\eta, \Lambda\right) \! \! \bigg)\Bigg]
\end{align}
and
\begin{align}
 f_{B_s} = \frac {1} {\sqrt {m_B}} \, \bar{f}_0^{ren}& \Bigg[1 +
     \alpha_1^{ren} (2 m_K^2 - m_\pi^2) + \alpha_2^{ren} (2 m_K^2 + m_\pi^2) \notag\\
 & \hspace*{-2pt} -  \, \frac {1+3g^2} {4 \, F_\phi^2} \bigg( 2 \, \mathbf{{I}^{ren}}
     \left(m_K, \Lambda\right) + \frac {2} {3} \, \mathbf{{I}^{ren}}
\left(m_\eta,\Lambda\right)\bigg)\Bigg].
\end{align}
Since at small mass, the residual integral
$\mathbf{I^{ren}} \left(m, \Lambda\right)$ tends to
$\mathbf{I^{d.r.}} \left(m\right)$,
the usual chiral expansion is recovered  at $m^2 << \Lambda^2$. At small
mass, the cut-off is just another way to regularize the calculation.

\section{The chiral extrapolation of $f_B$}

If we are going to use any meson loop calculation at larger masses in order to match
to the lattice, then all treatments are model dependent. We have argued above that
the use of chiral logs at these scales amounts to a bad model because it builds in
very large and spurious short distance effects. Our calculation above removes the
short distance effects in the one loop diagrams. This is then a reasonable formalism
to apply to the lattice calculation. The lattice calculation supplies the correct
short distance physics, described there through terms analytic in $m^2$ (linear
behavior, quadratic...). In addition, at smaller masses, our formulas naturally
include the chiral logarithms in the regions where they should be valid. This
motivates us to use the long-distance loop calculation in the chiral extrapolation
for B meson properties.

Let us first fit our expression to a caricature of the lattice data by matching the
data at two points. Such a linear extrapolation is appropriate for one loop since we
have only the constants and linear counterterms in the one loop expression. This fit
is demonstrated in Fig. 3, for various values of $\Lambda$. We see that the
extrapolation is smoother and that there is no large curvature induced at large
mass.
\begin{figure}[h]
\begin{center}
  \begin{minipage}[t]{.07\textwidth}
    \vspace{0pt}
    \centering
    \vspace*{55pt}
    \rotatebox{90}{$f_{B} \sqrt{m_B} \ [\mbox{GeV}^{3/2}]$}
  \end{minipage}%
  \begin{minipage}[t]{.93\textwidth}
    \vspace{0pt}
    \centering
    \includegraphics[width=0.99\textwidth,height=!]{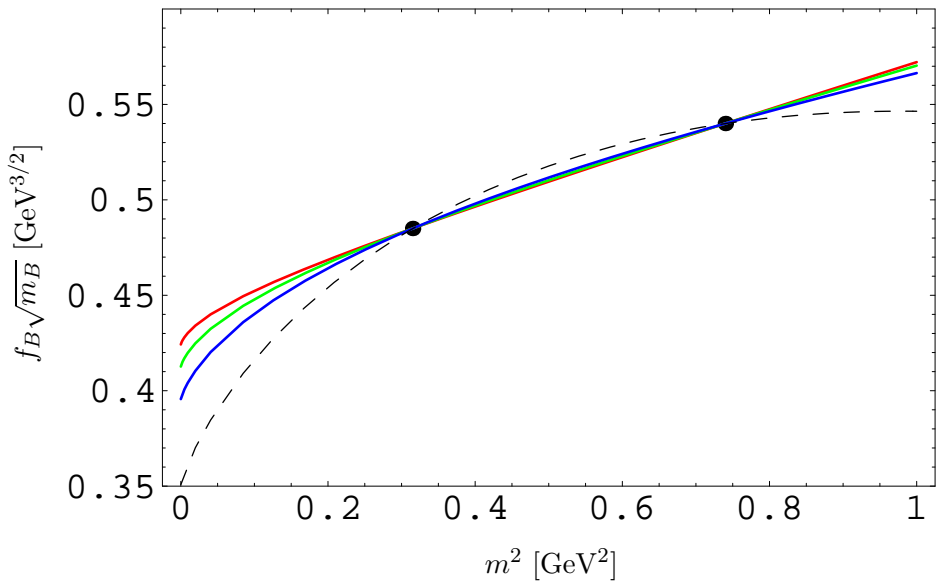}
    ${} \vspace*{-10pt}$
  \end{minipage}
\end{center}
$\hspace*{186pt} m^2 \ [\mbox{GeV}^2]$
\caption{$f_{B} \sqrt{m_B}$ as a function of $m^2$ fitted to the Lattice data points for
$\Lambda=400, 600, 1000$ MeV
and for the result from dimensional regularization (dashed)}
\end{figure}

There is a residual dependence of the extrapolated value on the parameter $\Lambda$.
This is shown in Fig. 4. In the range $\Lambda = 400 \mbox{ MeV} \rightarrow 1000
\mbox{ MeV}$, this amounts to a 5\% uncertainty in the extrapolated value. The
formula used in previous extrapolations corresponds to $\Lambda \rightarrow \infty$.
It is clear that the loop contributions that arise beyond the scale of $\Lambda =
1000$~MeV are of too short distance to be physically relevant for the effective
field theory - there is no reliable chiral physics beyond this scale.
\begin{figure}[h]
\begin{center}
  \begin{minipage}[t]{.07\textwidth}
    \vspace{0pt}
    \centering
    \vspace*{77pt}
    \rotatebox{90}{$f_{B}$ [GeV]}
  \end{minipage}%
  \begin{minipage}[t]{.93\textwidth}
    \vspace{0pt}
    \centering
    \includegraphics[width=0.99\textwidth,height=!]{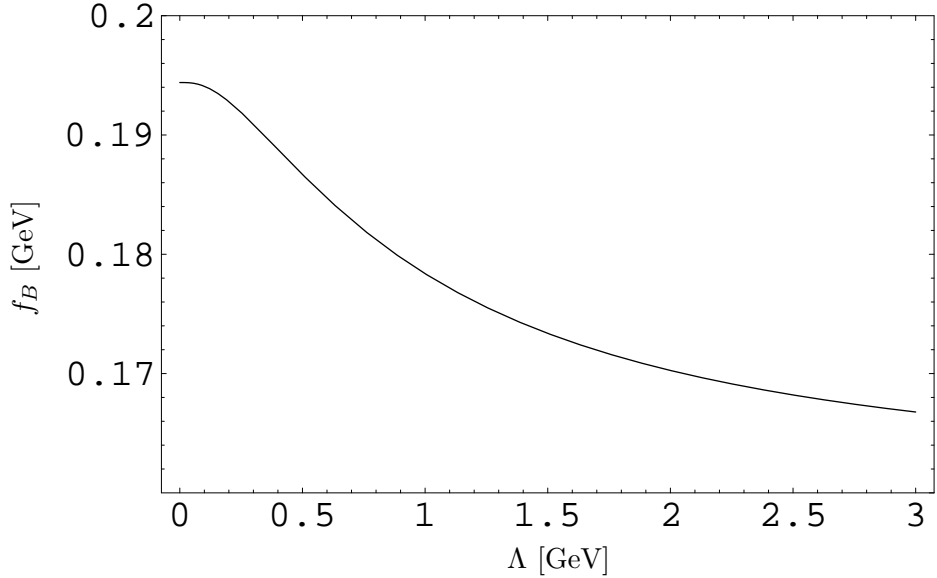}
    ${} \vspace*{-10pt}$
  \end{minipage}
\end{center}
$\hspace*{195pt}\Lambda$ [GeV]
\caption{$f_{B}$ at the physical pion mass as a function of $\Lambda$}
\end{figure}

This extrapolation can be systematically improved. Most favorably would be the
situation in which the lattice data can be calculated at smaller mass squared -
eventually no extrapolation would be needed. Even if the improved data goes only
part of the distance to the physical masses, it would remove some of the model
dependence of the result. The extrapolation needed would be smaller and the residual
$\Lambda$ dependence would be smaller. Another way that improvement possibly may be
made is with increased precision even at larger masses. As shown by Eq. 9 above, the
extrapolations for different $\Lambda$ values differ only at order $m^4/\Lambda^2$.
If one includes an extra $\mathcal{O}(m^4)$ in the one loop chiral calculation,
fitting to a quadratic expression, then the extrapolations will be in closer
agreement at this chiral order. Note however that the low mass region is still being
extrapolated by a one-loop chiral formula - this procedure is not equivalent to a
two-loop result in chiral perturbation theory.

As the lattice data reaches higher precision, it may be that the range of $\Lambda$
for which a good fit is obtained may shrink. While we are treating $\Lambda$ as a
regularization parameter, it is meant as a rough parameterization of a physical
effect - the transition from long-distance to short-distance in the loop
calculation. Therefore when using a fit to a given order in the chiral expansion,
the lattice data may only be describable with $\Lambda$ within some range near the
scale of this physical effect. Indeed, already the present data is a poor fit for
$\Lambda \to \infty$. Of course if one allows arbitrary orders in the chiral
expansion, with free parameters at each order, it is always possible to correct the
loop effect for any incorrect short distance behavior by adjusting the parameters.
However, when using the one loop integral with precise data it may not be possible
to obtain good fits for large values of $\Lambda$ without introducing {\it several}
new parameters at higher orders in the masses. In contrast, simpler fits with fewer
parameters may be obtained with $\Lambda$ within some optimal range.

Our procedure might be criticized as being a model, due to the choice of a
separation function and a separation scale. However, at large masses, the
dimensional regularization result is really more of a model as it introduces large
and unphysical short distance physics. Our procedure is the ``anti-model'' because
it removes most of that physics. The residual dependence on $\Lambda$ comes from the
ambiguity concerning how much of the short distance physics to remove. The value of
$\Lambda$ from the lattice results, introduced through the dipole cut-off,
parametrizes the short distance physics. However, this dependence can itself be
adjusted by using the coefficients of the chiral lagrangian. Despite the decoupling
of the loop at large mass, we retain all of the correct chiral behavior in the limit
of small quark mass.

\section{Application to $B_B$}

 All of the preceding formalism can also be applied to the chiral extrapolation of
the $B_B$ parameter for $B-\bar{B}$ mixing. We have reproduced the calculations of
Ref. \cite{grinstein, sharpe} using throughout the method of long distance
regularization. As above, only the integral $\mathbf{I^{ren}}$ is needed in the
final answer. The chiral formulas after renormalization of the parameters are
\begin{align}
 B_{B_d} =  B_0^{ren} \, & \Bigg[1 + \beta_{1}^{ren} \, m_\pi^2 + \beta_{2}^{ren} \, (2
m_k^2 + m_\pi^2) \notag\\
 & -  \, \frac {1-3g^2} {4 \, F_\phi^2} \bigg(\mathbf{I^{ren}} \left(m_\pi,
\Lambda\right)
   + \frac {1} {3} \, \mathbf{I^{ren}} \left(m_\eta, \Lambda\right) \! \! \bigg)\Bigg]\\
\notag \\
 B_{B_s} = B_0^{ren} \, & \Bigg[1 + \beta_{1}^{ren} \, (2 m_K^2 - m_\pi^2) +
\beta_{2}^{ren} \, (2 m_k^2 + m_\pi^2) \notag\\
 & -  \, \frac {1-3g^2} {3 \, F_\phi^2} \, \mathbf{I^{ren}} \left(m_\eta,
\Lambda\right)\Bigg]
\end{align}
in the same notation as before. Here the new chiral constants $B_0$, $\beta_1$,
$\beta_2$ describe the intercept and slope of the chiral expansion. At small masses
the usual dimensional regularization results of Ref.~\cite{grinstein, sharpe} are
recovered in the limit of small $m/\Lambda $, as is seen using    Eq.~9.

The chiral corrections for $B_B$ are proportional to $1-3g^2$, while in the case of
$f_B$ the corrections contain the factor $1+3g^2$. This modification makes an
important change in the result. For the coupling $g = 0.59$ that is favored by
recent measurements \cite{couplingg} and supported by recent lattice calculations
and theoretical predictions \cite{couplingg}, the factor $1-3g^2$ almost vanishes.
In this case, the one loop chiral corrections are tiny whether one employs the
standard scheme or our long-distance regularization methods. (See also
\cite{Aoki:2002bh} for a discussion of this effect). For this reason, we do not
display the numerical effect of the chiral extrapolation of $B_B$. Use of a
significantly smaller value of the coupling $g$ would lead to measurable effect in
the $B_B$ extrapolation.

\section{Conclusions}

We have presented a method for the extrapolation of lattice data to smaller quark
masses. This includes the chiral logarithm in the region where it is valid. It has
the advantage that it removes the large and unphysical short distance effects that
caused problems in previous methods. There is still some residual model dependence
that is visible in the variation of the results on $\Lambda$. However the
extrapolations are better behaved than previous ones. The residual uncertainty in a
linear extrapolation (i.e. with a slope proportional to $m^2$ and no chiral
logarithm) for $f_B$ is about 5\% when the cut-off is constrained to the range 400
MeV--1000 MeV. For $B_B$ the uncertainty in the chiral extrapolation is negligible
for $g=0.59$. We would recommend that our method only be applied for values in this
range.

The chiral corrections have the effect of producing a slight decrease in the
extrapolated values of $f_B$ and $B_B$ when compared to an extrapolation which does
not include chiral effects. This is the effect of the non-analytic behavior of the
chiral logarithm at long distance. Our estimates suggest that the decrease due to
the chiral log puts the chirally corrected result at $0.945 \pm 0.025$ of the
uncorrected extrapolation for $f_B$. We hope that our method will be applied in
future extrapolations of lattice data.

\section*{Acknowledgement} We thank Eugene Golowich, Barry Holstein, Damir Becirevic
and Laurent Lellouch for useful conversations. This work was supported in part by
the US National Science Foundation and in part by TMR, EC Contract No. ERB
FMRX-CT98-0169, by MCYT (Spain) under grant FPA-2001-3031 and by ERDF funds from the
European Commission. JFD thanks the Institute des Hautes \'{E}tudes Scientifiques
for its hospitality during the writing of this manuscript.

\end{document}